\documentclass[preprint,aps,nofootinbib]{revtex4}

\usepackage{graphicx}
\usepackage{dcolumn}
\usepackage{bm}
\usepackage{amssymb}
\newcommand{\doci}{$^\circ$C}

\newcommand{\tsan}{$\tau_3$}

\begin{document}   
\title{Measurement of positron lifetime to probe the mixed molecular states
 of liquid water} 

\author{Katsushige Kotera}
 \affiliation{Physics Department, Osaka University,
   Machikaneyama 1-1, Toyonaka, Osaka 560-0043, Japan}
\author{Tadashi Saito}
\affiliation{Radioisotope Research Center, Osaka University,
Machikaneyama 1-1, Toyonaka, Osaka 560-0043, Japan}

\author{Taku Yamanaka}
\affiliation{Physics Department, Osaka University,
Machikaneyama 1-1, Toyonaka, Osaka 560-0043, Japan}

\begin{abstract}	
Positron lifetime spectra were measured in liquid water at temperatures 
between 0\doci\ and 50\doci.
The long lifetime of $ortho$-positronium atoms ($o$-Ps) determined by electron 
pick-off in molecular substances decreases smoothly by 10\% as the temperature
is raised.
This lifetime temperature dependence can be explained by combining 
the Ps-bubble model and the mixture state model of liquid water.
\end{abstract}

\maketitle
\section{\label{sec:level1}Introduction}

Water has been studied extensively because it is a fundamental liquid with 
unique properties. 
For example, its boiling point under atmospheric pressure, 100\doci, 
is unusually high for its molecular weight, 18.
In addition, its volume decreases as it melts, unlike other liquids. 
Moreover, it has an anomalous density maximum at 4\doci.
These characteristics have been studied in detail using various methods. 
Consequently, many models have been suggested.
Nevertheless, no definite mechanism has been established to explain these 
anomalous behaviors of water, even though each model can explain some specific 
phenomena.

One sensitive method to investigate molecular states in solid and liquid forms
is the positron annihilation lifetime (PALT) method.
When positrons are injected into a liquid,
some of them form positroniums with the surrounding electrons.
The positronium lifetime is determined by the rate of annihilation 
between the positron in the positronium and surrounding electrons, 
also known as pick-off.
As described later, according to the Ps-bubble model 
\cite{Brandt60,Brandt66,Tao72},  
the lifetime should be longer at higher temperatures in liquid water.
However, 
V\'{e}rtes $et\ al.$ \cite{Vertes} showed that the positronium lifetime has 
a general decreasing trend for higher temperature, although it oscillates 
between 38\doci\ to 60\doci. 

We measured PALT in liquid water between 0\doci\  and 50\doci.
This paper explains that the positron lifetime decreases smoothly
as the temperature rises. This behavior is explainable by
combining the Ps-bubble model and a two-state mixture model of water.

\section{Positron annihilation lifetime method and the Ps-bubble model}

This section explains PALT and the Ps-bubble model.

Positrons injected in material can undergo three different processes, 
each with a different lifetime.
The shortest lifetime ($\tau_1$) occurs by a formation and annihilation of 
  $para$-positronium ($p$-Ps) in which the spins are anti-parallel.
In that situation, the electron forming $p$-Ps with the positron is captured from 
 the surrounding molecules.
The lifetime of $p$-Ps in vacuum, $\sim$ 0.1 ns, 		
is too short to be affected by surrounding matter.
 
  The second lifetime ($\tau_2$) is caused by positron annihilation without 
forming a bound state.
 
 The third lifetime ($\tau_3$) results from formation of an 
$ortho$-positronium ($o$-Ps)  with parallel spins. 
  
  The $o$-Ps has an intrinsic lifetime of 142 ns in vacuum.  
  However, in material, the lifetime is typically several nanoseconds 
because it  annihilates with an electron in the surrounding molecules. 
 As a result of this pick-off process, $\tau_3$ is sensitive to 
the electron-state of the  surrounding substance. 
In liquid, $o$-Ps is generally considered to push out the surrounding 
molecules and form a Ps-bubble.
 The wave function of $o$-Ps exudes from the surface of Ps-bubble, 
which comprises electrons from surrounding substances. 
The overlap of $o$-Ps wave function and  
an electron wave function of the surroundings determines the pick-off rate. 
In the Ps-bubble model, the wave function of $o$-Ps is trapped inside a 
spherical infinite-depth well potential whose radius is the sum of 
the radius of the Ps-bubble and the exuding depth.
Therefore, \tsan\ is a function of the Ps-bubble radius 
 \cite{Tao72,Eldrup82,MogensenBook}.
 
 Nakanishi $et\ al.$ introduced a semi-empirical correlation between 
$\tau_3$ and  the Ps-bubble radius \cite{Nakanishi88} as:
\begin{equation}\label{N-J}						
 \tau_3 = \biggl[ 2\biggl\{ 1- \frac{R}{R+\Delta R} + \frac{1}{2 \pi} 
   \sin \biggl( \frac{2 \pi R}{R+\Delta R} \biggr) \biggr\} \biggr]^{-1}, 
 \end{equation} 
 where \tsan\ is measured in nanoseconds, 
$R$ is the Ps-bubble radius, and $\Delta R$ is the exuding depth of the $o$-Ps 
wave function into the surrounding electron wave functions. 
In this model, $R$ is determined by the balance  
between the zero-point energy of $o$-Ps and the surface tension of 
the surrounding substance.
The balance is represented as 
\begin{equation}\label{0} 
  \frac{\partial}{\partial R}(E+4 \pi R^2 \gamma )=0, 
\end{equation} 
  where $E$ is the zero point energy of $o$-Ps and $\gamma$ is the surface  
tension of bulk water. 
$E$ is expressed as 
 \begin{equation}\label{E}  
 E \sim \frac{\hbar^2 k^2}{4 m_e} = \frac{\hbar^2 \pi^2}{4 m_e (R+\Delta R)^2}, 
 \end{equation}
where $m_e$ is the electron mass and $k$ is the momentum of $o$-Ps. 

For the case of water, $\gamma$ decreases as the temperature rises;  
thereby the bubble becomes larger. 
By this argument alone, $\tau_3$ is longer for higher temperatures.  

\section{Experiment} 
\subsection{Measurement of positron lifetime}			 
We employed the positron annihilation lifetime technique using $ ^{22}$Na 
as a positron source. 
$ ^{22}$Na is put in the sample center.  
Positrons emitted into the sample are annihilated 
through the process mentioned in the previous section. 
When $ ^{22}$Na emits a positron, it simultaneously emits a 1275 keV photon. 
When a positron is annihilated with an electron, two 511 keV photons are 
emitted. 
In light of those facts, we measured the time difference between the emission 
of 1275 keV photon and the emission of two 511 keV photons.

\begin{figure}										
\begin{center}\includegraphics[width=8cm]{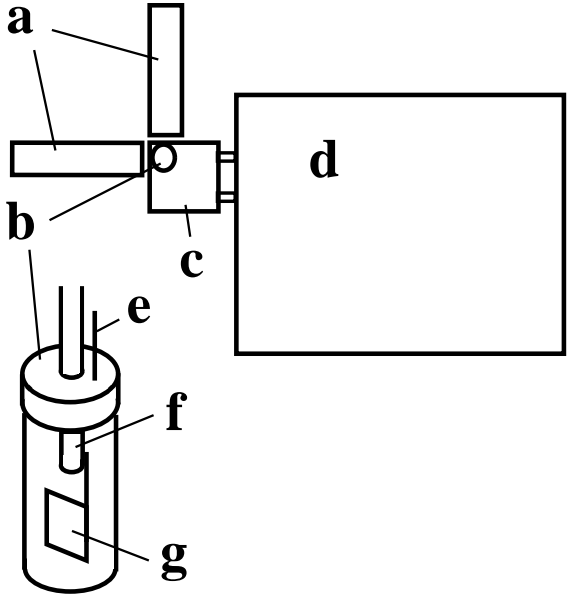}		
\caption{\label{fig:apparatus}A schematic of the apparatus. 	
	a. $\mbox{BaF}_2$ scintillators with photomultiplier tubes, 	
	b. Sample vial, c. Satellite water bath	
	(60 mm $\times$ 60 mm $\times$ 100 mm (height)), 	
	d. Main water bath (300 mm $\times$ 500 mm $\times$ 160 mm (height)),	
	e. Needle for pressure equilibrium, f. K-thermocouple, 	
	g. Positron source.	
	}
\end{center} 
\end{figure}

We used the apparatus shown in Fig.~\ref{fig:apparatus}.	
Powder of 1.6 MBq $\ ^{22}$NaCl wrapped between 8 $\mu$m thick 
polyimide films was used as a positron source. 
The polyimide film is known to have no  
\tsan\ component \cite{polyimide}. 
The source was held inside a 20 mm diameter 25 ml glass vial. 
The vial was then filled with de-gassed distilled water. 
We estimate that 4.4\% of the positrons were absorbed in polyimide film;  
more than 95.5\% of the positrons are absorbed within 0.5 mm of water 
\footnote
	{
	Using the linear absorption coefficient($\alpha$),
	$\alpha = 4 d/E^{1.6}_{\mbox{max}}(\mbox{cm}^{-1})$, where $d$ 
	is the density of absorber and $E^{1.6}_{\mbox{max}}$ 
	is the maximum energy of a positron in MeV.
	}. 
The vial was submerged in a water bath to maintain the sample 
water temperature within $\pm0.1$\doci. 
The sample water pressure was maintained at 1 atm by inserting a 0.9 mm  
 inner diameter needle into the vial. 
 
The 1275 keV photon (start signal) was detected using a barium fluoride  
 ($\mbox{BaF}_2$) scintillator (38.1 mm diameter, 25.4 mm thickness)  
with a photomultiplier tube (H3378-51; Hamamatsu Photonics, K.K.) and  
 one 511 keV photon (stop signal)  
was detected by another $\mbox{BaF}_2$ detector. 
The two $\mbox{BaF}_2$ detectors viewed the vial at a 90$^\circ$ 
opening angle;  
they were held 27 mm away  
 from the center of the sample water. 
 
 The time difference between the start and stop signals was measured with 
 a time-to-amplitude convertor and a multi-channel analyzer. 
 The time resolution of this system is $\sim280$ ps. 
 We measured the positron lifetime at 14 temperatures between 0\doci\ and 
50\doci. 
In most cases, we performed four runs at each temperature point. We used 
a new water sample for each run.  
Each run collected one million events in 1500 s. 

\subsection{Analysis}							
\begin{figure}									
\begin{center} 
\includegraphics[width=8cm]{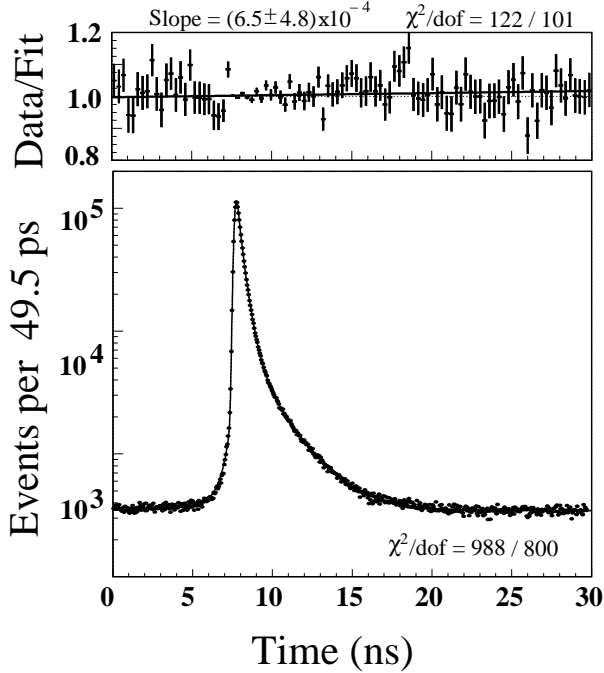}		
\caption{\label{fig:spectrum} A typical positron lifetime spectrum in liquid 
         water (bottom) and the data-to-fit ratios (top). 
	Data are shown in dots. A fitting result by MINUIT is shown 
	in a solid line in the spectrum.} 
\end{center} 
\end{figure}

A typical lifetime spectrum acquired in our experiment is shown in  
Fig.~\ref{fig:spectrum}.	

To extract lifetimes, the spectrum was fitted for a model function, 
\begin{equation}\label{fit}  
  S(t) = \sum_{i=1}^3 \frac{I_i}{\tau_i} \int\mbox{e}^{-t'/\tau_i} 
  g(t-t') \theta(t'-t_0)dt' +
  k  \sum_{i=1}^3  \frac{I_i}{\tau_i} \int\mbox{e}^{t'/\tau_i} 
  g(t-t') \theta(t_0 -t')dt' + C, 
\end{equation} 
where $i$ is the index for different lifetimes, 
$I_i$ is intensity, \ $\tau_i$ is the lifetime,  
and $g(t-t')$ is the Gaussian describing the time resolution.  
Also, $\theta(t)$ represents a step function in which 
$\theta(t)=1\ (t\ge 0)$ and $0\ (t < 0)$, $t_0$ is the delayed timing of 
the start signal,  
 and $C$ is a constant background rate. 
 The second term is added to accommodate events in which a 511 keV photon gave 
 the start signal and a 1275 keV photon gave the stop signal. 
 Ten parameters,\ $I_{1,2,3}, \tau_{1,2,3}, t_0, k, C$, and the 
 sigma of the Gaussian were fitted using MINUIT \cite{MINUIT} fitting code, 
which is commonly used in analyses of high-energy physics. 
A fitting result achieved using MINUIT is shown as a solid line 
in Fig. \ref{fig:spectrum}. 

We used POSITRONFIT \cite{posfit},
which is a popular code in positron lifetime analysis, only to crosscheck 
our results because 
POSITRONFIT requires the time resolution as an input parameter. 
The results are sensitive to the lower boundary of the fitting range.
Both codes gave mutually consistent results. 
The mean value of the difference between both codes taken over all 
measurements was 0.008 ns; their standard deviation was 0.016 ns. 

\section{Results}		
\begin{figure}								
\begin{center}									
\includegraphics[width=8.0cm]{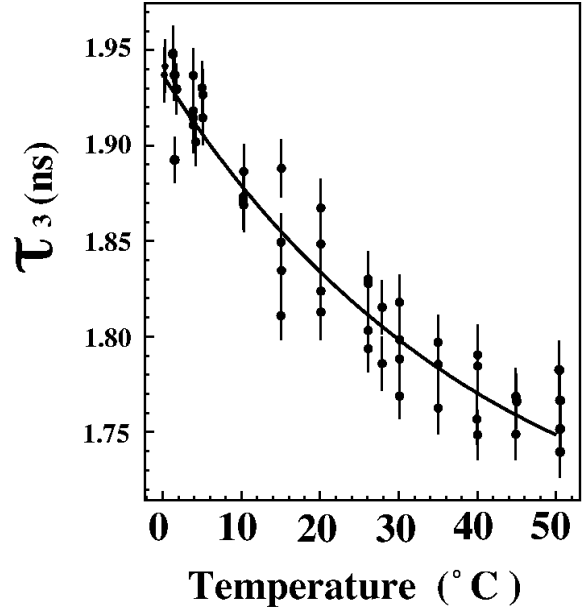} 		
\caption{\label{fig:tau3_fit}$\tau_3$ in water as a function of 
			temperature.
			Dots show data. Vertical lines show errors.
			The solid line shows a fitting result by Ps-bubble 
			model combined with the two-state model.}	 
\end{center}
\end{figure}

Temperature dependence of $\tau_3$ is shown in Fig. \ref{fig:tau3_fit}.
The $\tau_3$ decreases smoothly by 10\% as the temperature is raised 
from 0\doci\ to 50\doci. 
Our \tsan\ at 20\doci, $1.839\pm0.015$ ns,
agrees with the measurement of Mogensen, 1.85 ns \cite{Mogensen_water}.

\begin{figure}								
\begin{center}
	\includegraphics[width=5.0cm]	{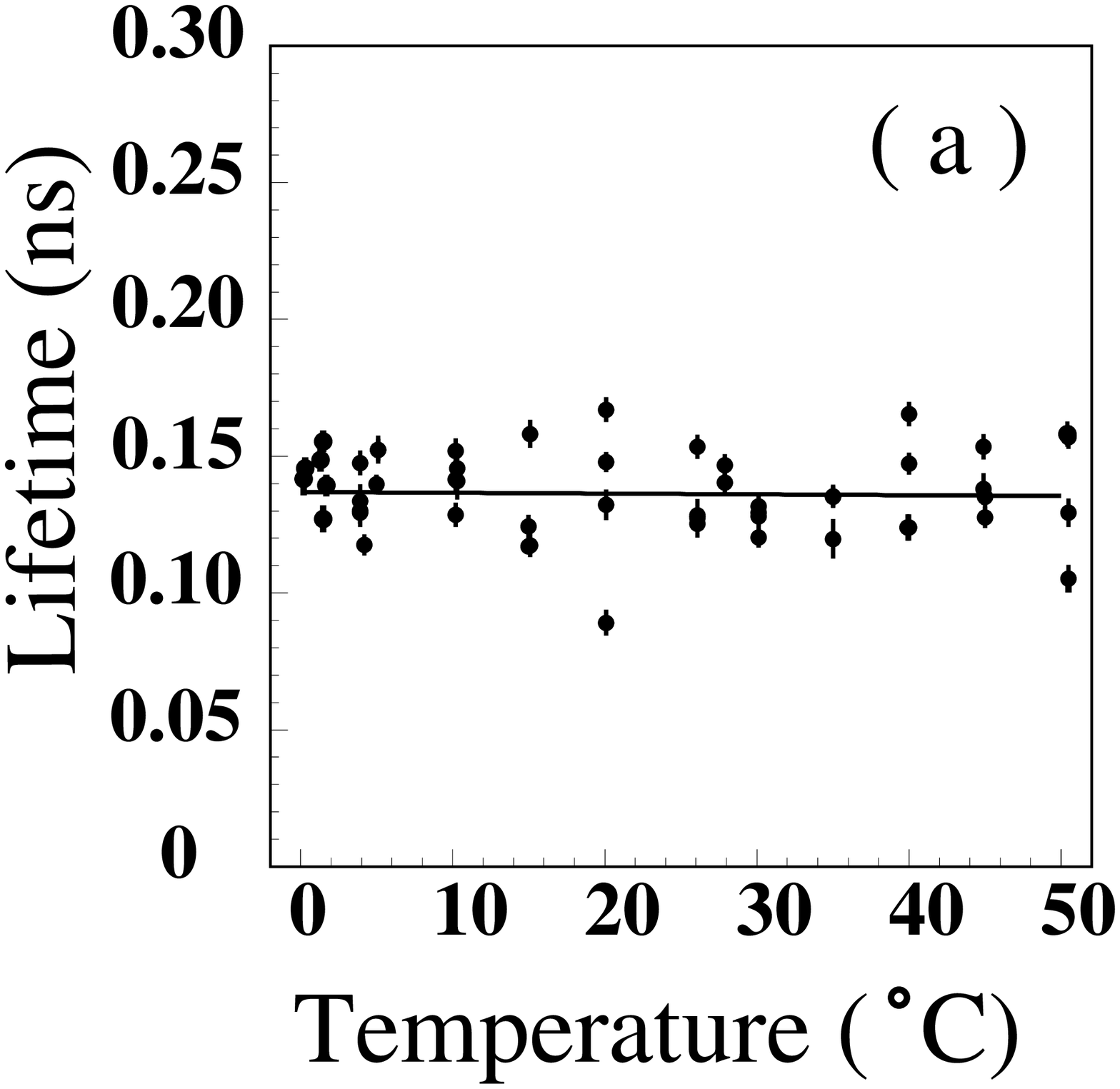}	
	\includegraphics[width=5.0cm]	{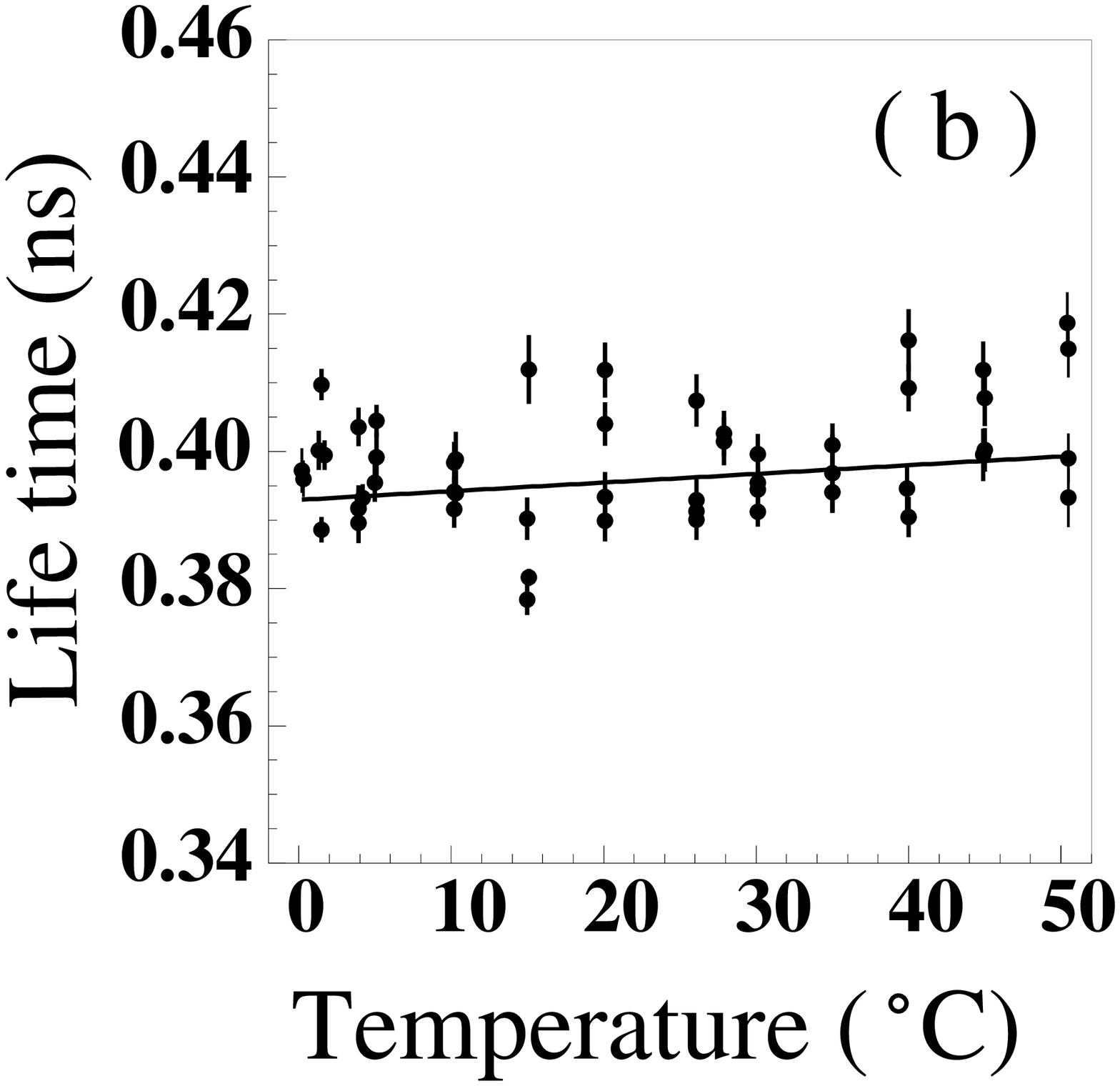}	
	\includegraphics[width=5.0cm]	{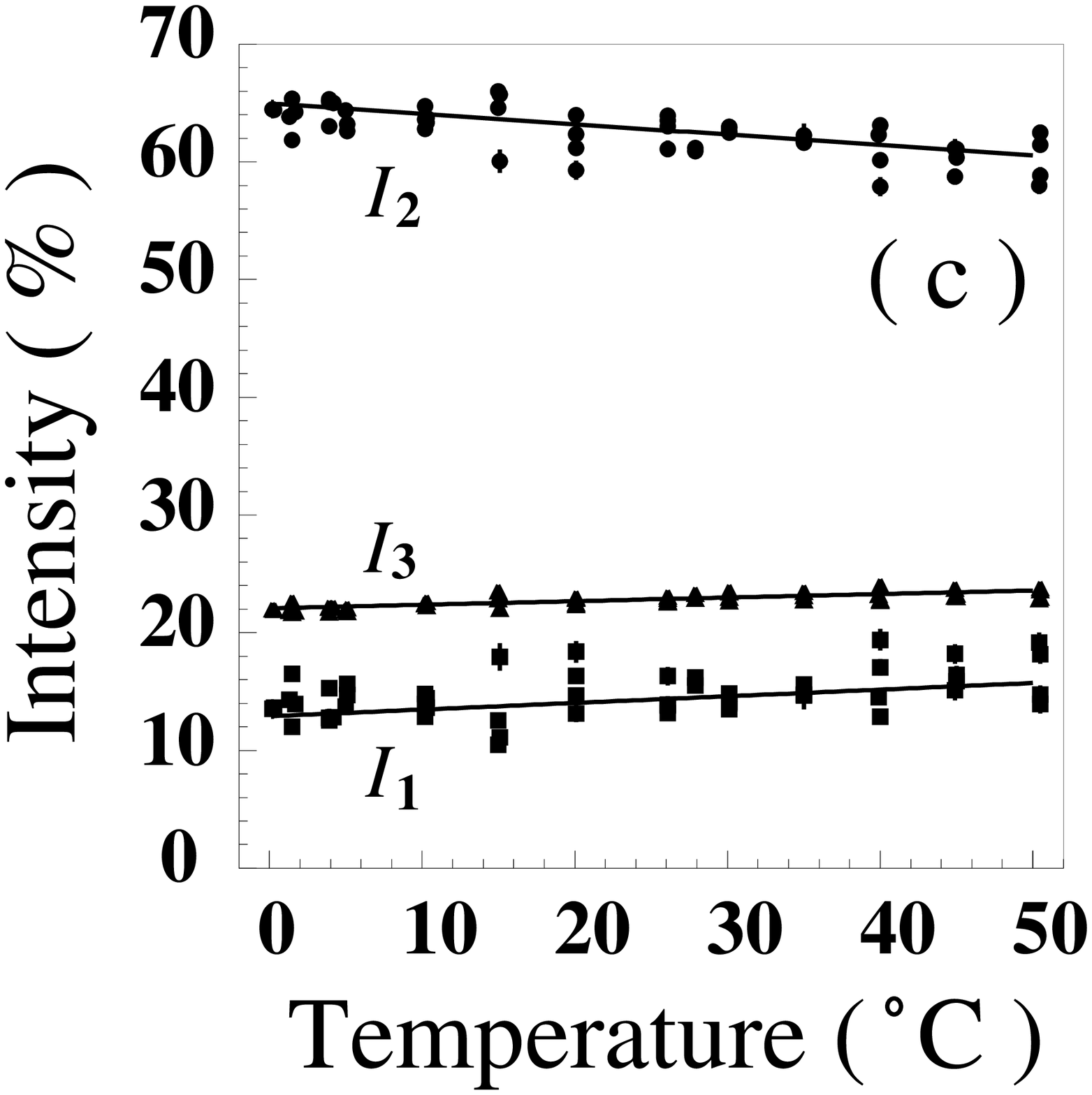}	
 \caption{\label{fig:tau_all}
          Temperature dependencies of lifetimes, $\tau_1$ and $\tau_2$
	in water and intensities of three lifetime components.
	Dots show data and vertical lines indicate the fitting error. 
	Solid lines are results of linear regression.			
	a, b: lifetime component $\tau_1, \tau_2$, respectively,
	c: intensities of three components.}
\end{center}
\end{figure}

\begin{figure}[width=11cm]
\begin{center}

	\includegraphics[width=3.5cm]	
	{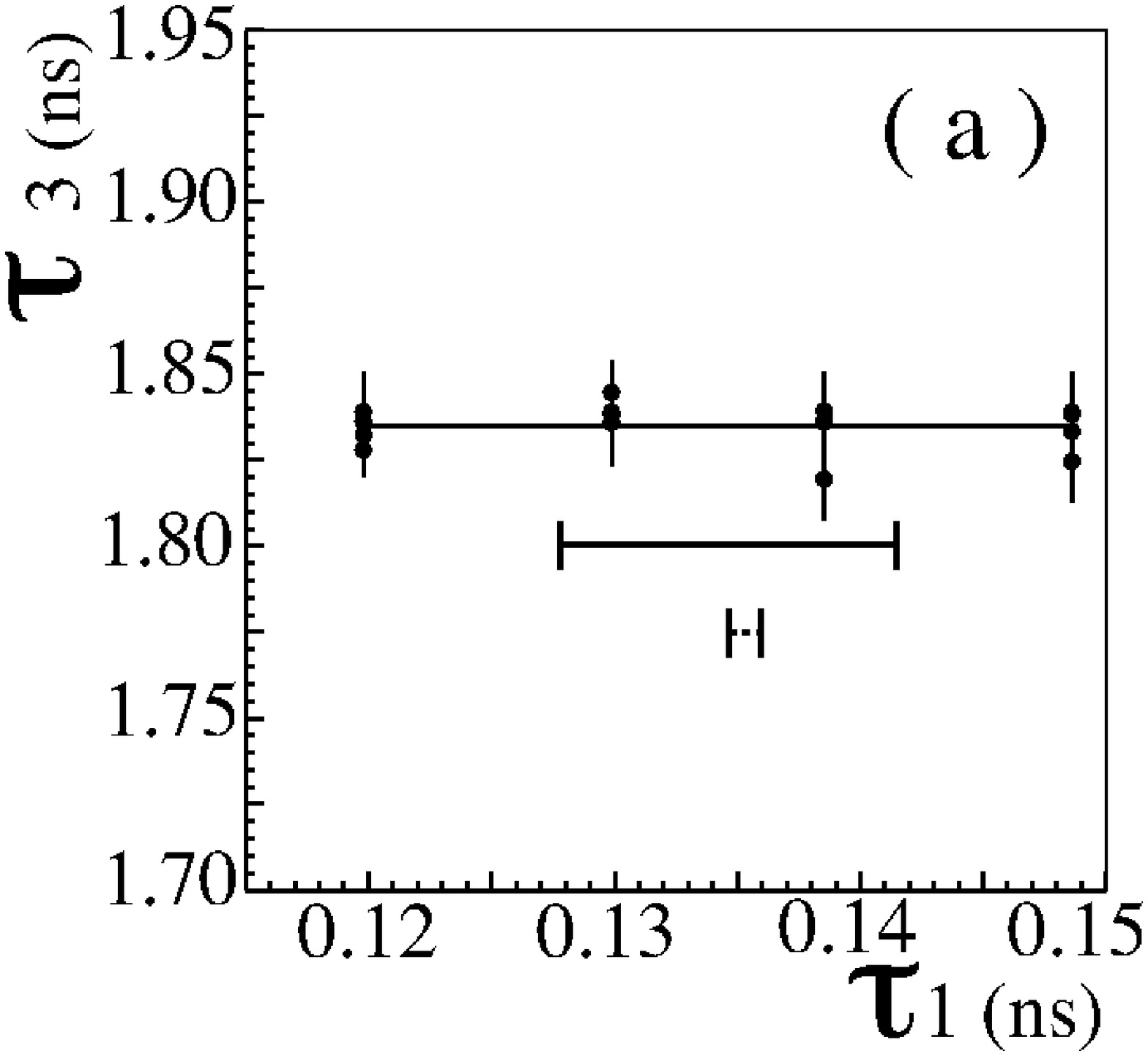}				
	\includegraphics[width=3.5cm]	
	{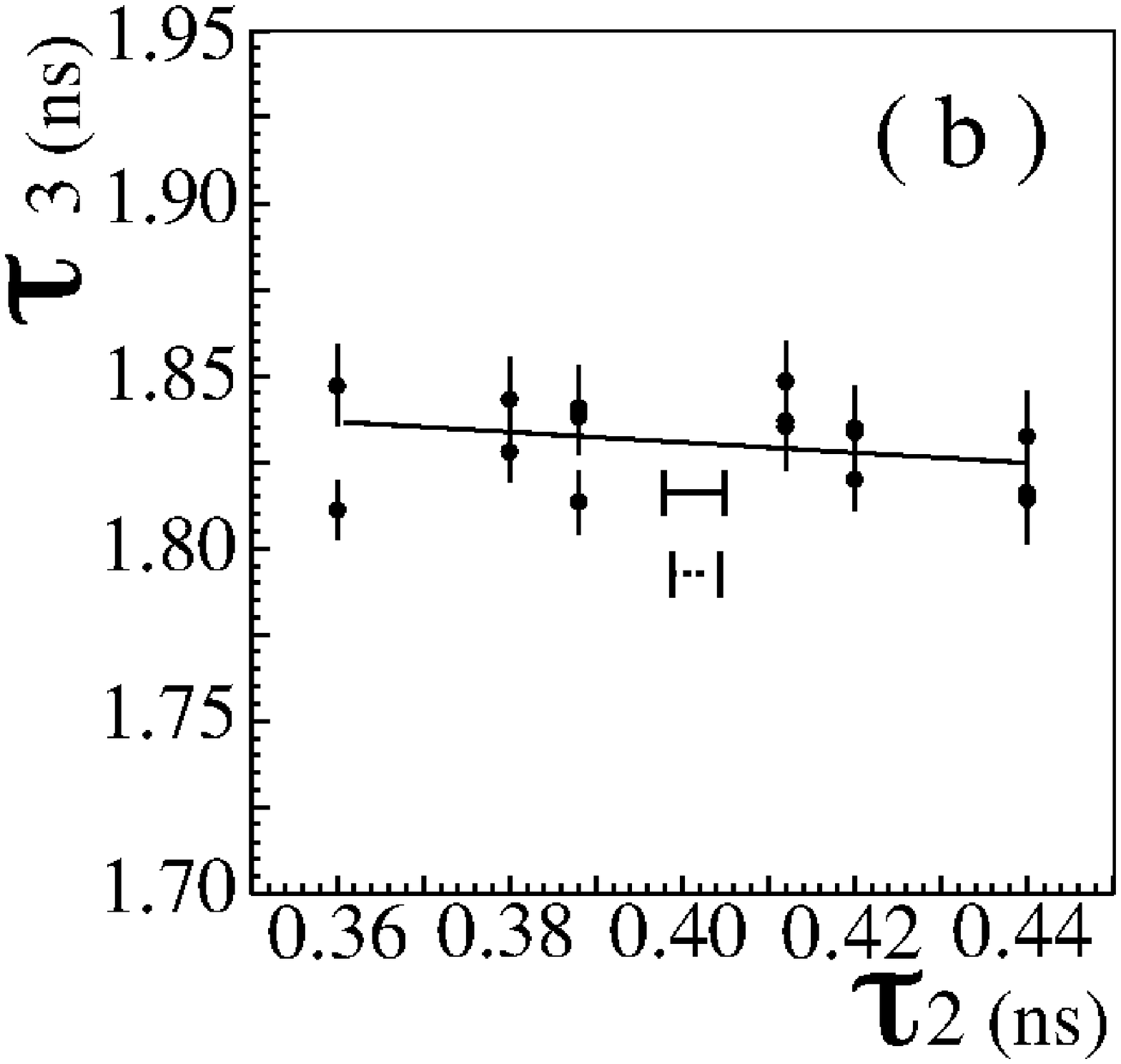}				
	\includegraphics[width=3.5cm]
	{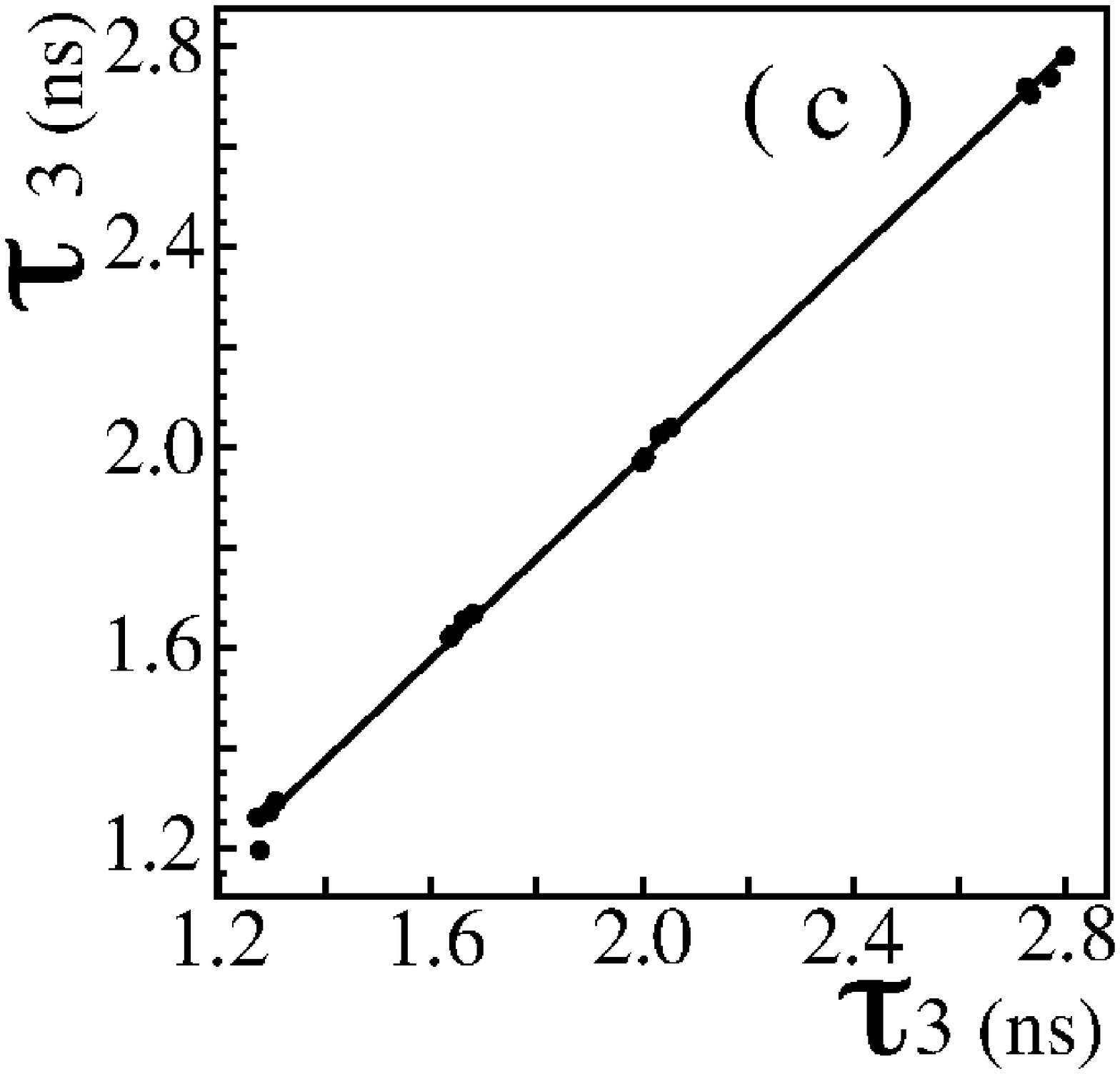}				
	
	\includegraphics[width=3.5cm]
	{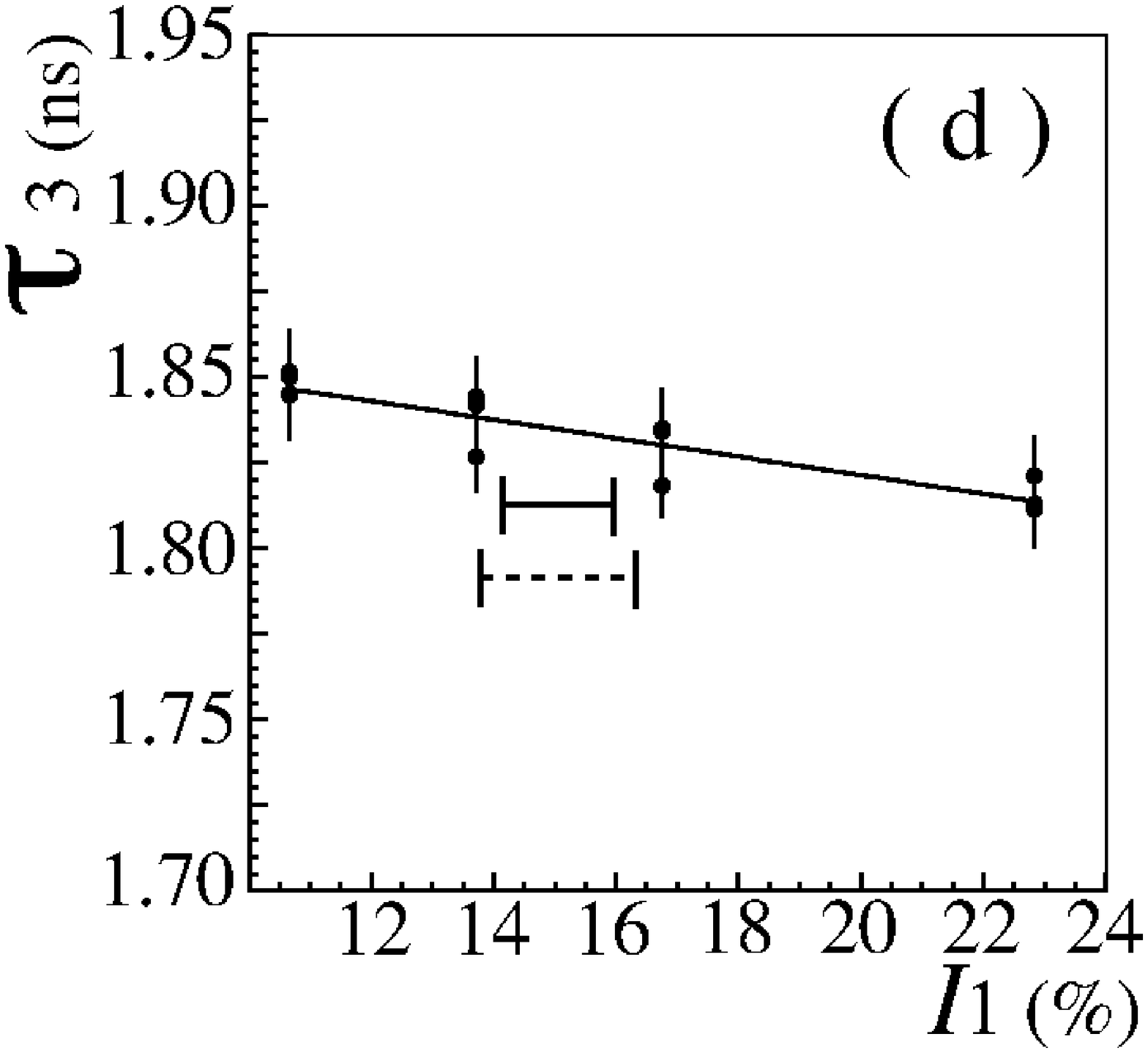}				
	\includegraphics[width=3.5cm]	
	{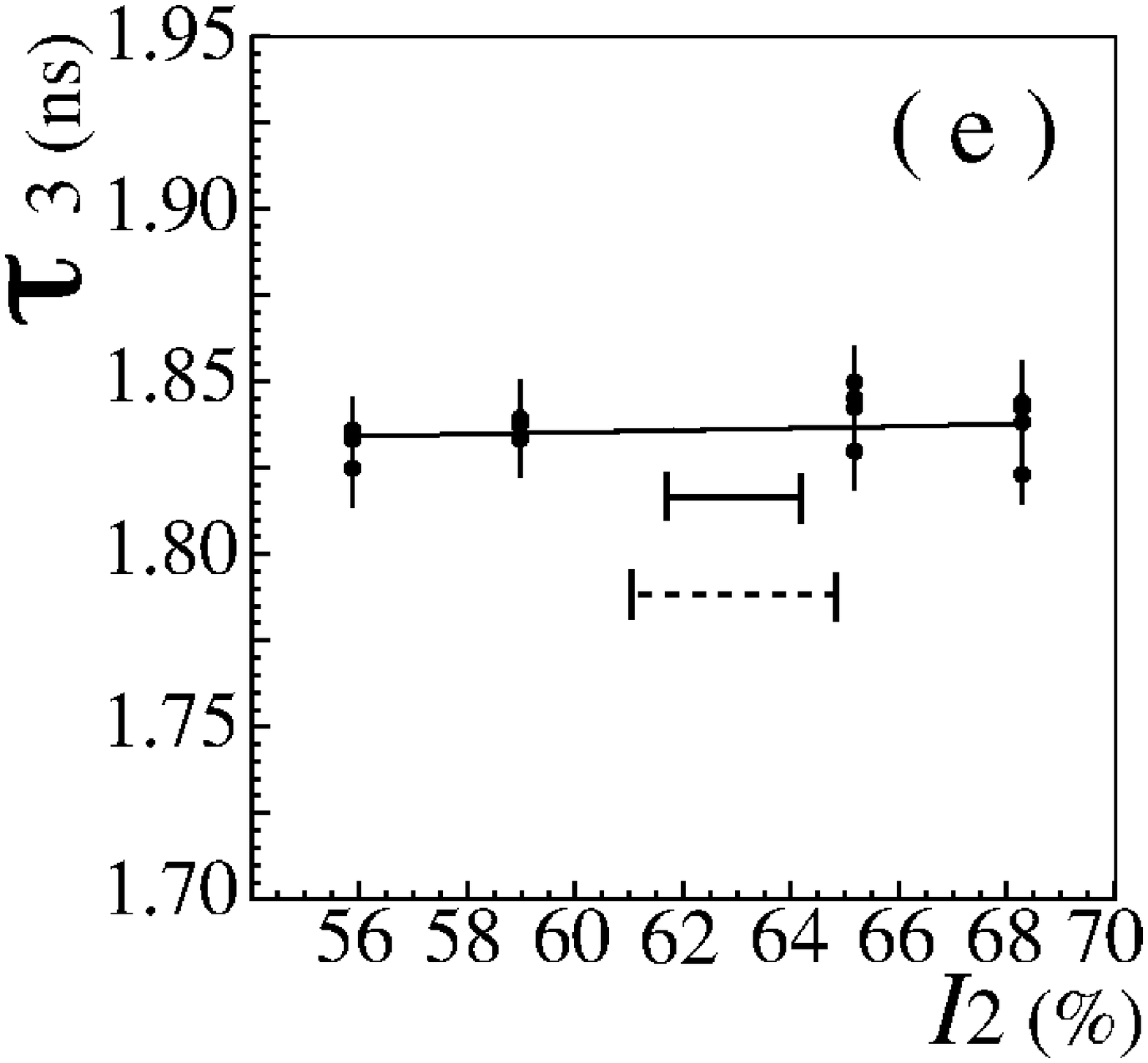}				
	\includegraphics[width=3.5cm]
	{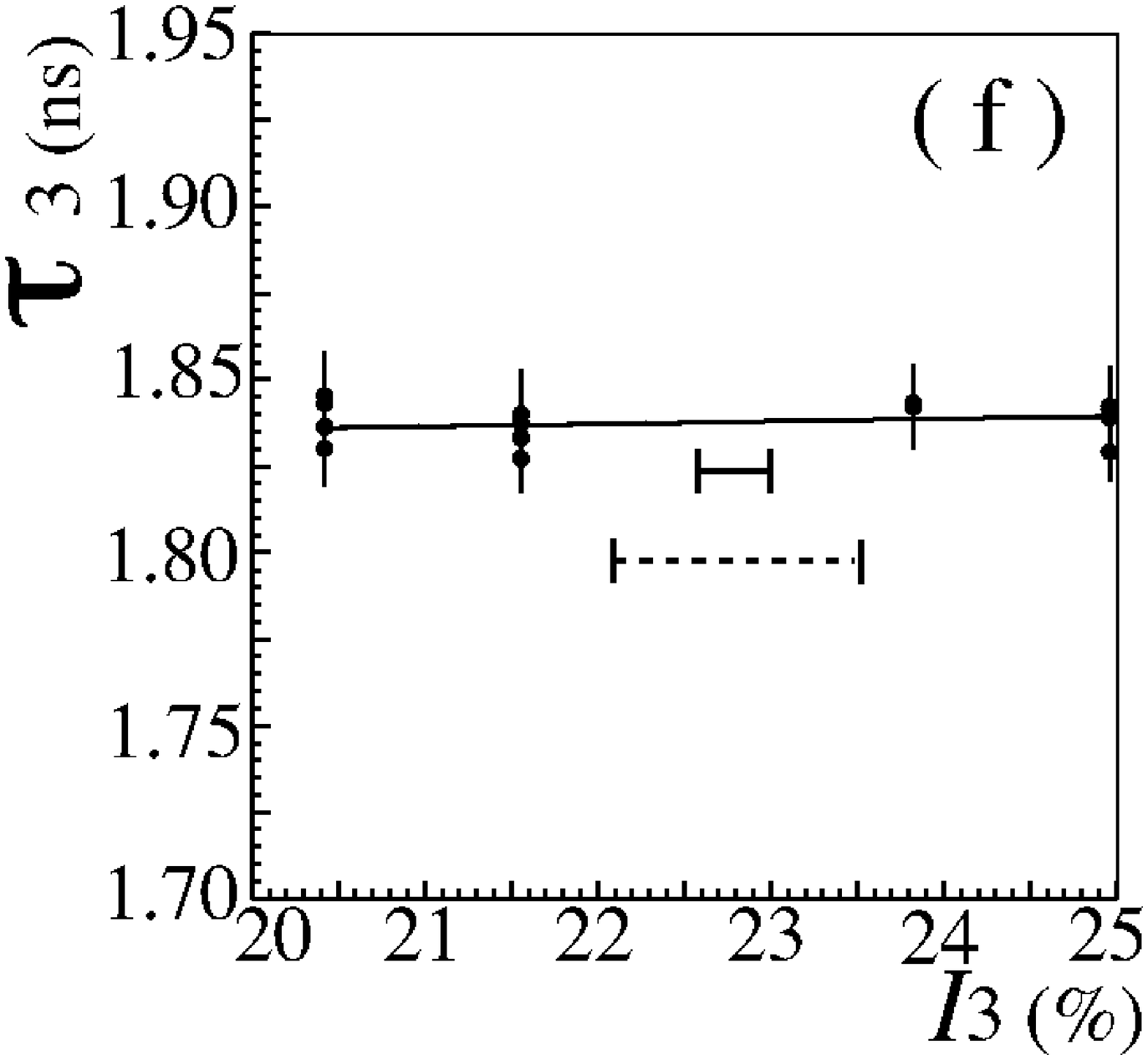}				
	
\caption{\label{influences}Fitted $\tau_3$ as a function of other 
			parameters studied with artificial data samples:
			 (a) $\tau_1$, (b) $\tau_2$, (c) $\tau_3$,	
			 (d) $I_1$, (e) $I_2$, and (f) $I_3$.			
			 Horizontal solid bars show averages of 
			 the standard deviations of parameters 			
			 at each temperature point;			
			 dashed bars show the maximum changes 			
			 of parameters			
			 between 0 and 50\doci.}
\end{center}
\end{figure}

Behaviors of other lifetime parameters are shown in Fig. \ref{fig:tau_all}.
We confirmed that the behavior of \tsan\ is not caused by a change in other 
lifetime parameters.
To do so, we produced artificial spectra that represent a sum of three 
exponentials, $\sum_{i=1}^3I_i\exp (t/\tau_i)/\tau_i$, 
convoluted by a Gaussian time resolution. 
The number of events in each 0.0495 ns time bin was made to follow 
a Poisson distribution.
We produced such artificial data samples with different $\tau_i$ and 
$I_i$, and fitted for them. 
Figure \ref{influences} shows the fitted $\tau_3$ as a function of 
the varied parameters. 
Solid lines show the linear fit to the points.
First, the fitted $\tau_3$ agrees with the input $\tau_3$ within 
$0.01\pm0.01$ ns, as shown in Fig. \ref{influences}(c). 
Second, the fitted $\tau_3$ does not depend on $\tau_1$, $\tau_2$, $I_2$, 
and $I_3$.
Horizontal bars show the variations and errors 
of parameters in the measured temperature range.
The maximum deviation of \tsan\ is calculated by multiplying the slope of 
the linear fit and the quadratic sum of the variation and the error of 
the other parameters. Deviations resulting from $\tau_1$, $\tau_2$, $I_2$, 
and $I_3$ are  $0.000\pm0.004$ ns, $0.0014\pm0.0008$ ns,  
$0.0015\pm0.0029$ ns, and $0.001\pm0.003$ ns, respectively.
The only apparent dependence is on $I_1$, as shown in 
Fig. \ref{influences}(d);
its effect on $\tau_3$ is $0.009\pm0.002$ ns.
The total effect on $\tau_3$ that is attributable to the error and swing 
of other parameters in the temperature range is 0.011 ns.
Consequently, we can conclude that the temperature dependence of \tsan\ 
is not caused by changes in other lifetime parameters.

\section{Discussion}
The behavior of our $\tau_3$ cannot be explained by the original 
Ps-bubble model.
For that reason, we apply the two-state mixture model to the Ps-bubble model.
Vedamuthu $et\ al.$ showed that the two-state mixture model 
can reproduce density of $\mbox H_2 \mbox O$ in 
the temperature range between -30 and +70\doci\ with very high precision 
 \cite{Vedamuthu94}.
In this model, liquid water comprises two dynamically inter-converting 
mixed micro-domains whose bonding characteristics
are similar to those of ice I$h$ (lower density) and ice II (higher density) 
\cite{Urquidi PRL 99}.
The total specific volume of liquid water is given as
\begin{equation} \label{Va}
  V(T,P) =   [1-f_{\mbox {II}}(T,P)]V_{\mbox I}(T,P) + 
  f_{\mbox {II}}(T,P)V_{\mbox  {II}}(T,P),
\end{equation} 
where I (II) indicates lower (higher) density bonding, $f_{\mbox {II}}$ 
is the mass fraction of the higher density bonding type, 
and $V_{\mbox{I(II)}}$ is the specific volume of
the lower (higher) density type bond.

We modified the Ps-bubble model by introducing the two-state mixture model 
with the following assumptions.
\begin{enumerate}
    \item Positronium forms a bubble (Ps-bubble) with a radius $R$ in water
    	as the original Ps-bubble model.
    \item Water consists of two different molecular bond types, I and II; 
       	the ratio  $f_{\mbox {II}}$ is a function of the temperature as
	given by Vedamuthu $et\ al$. \cite{Vedamuthu94}.   
     \item The pick-off rate is determined by an overlap between the    	
     $o$-Ps wave function and the electron wave function of the water;    	
     it can be parameterized by exuding depths,   	
      $\Delta R_{\mbox {I}}$ and $\Delta R_{\mbox {II}}$	
      for each bonding type.    
     \item The bubble radius, $R$, is given by a macroscopic surface    	
    tension $\gamma$, which is determined by the temperature \cite{gamma},  
	 $\Delta R_{\mbox {I}}$ and $\Delta R_{\mbox {II}}$ .
\end{enumerate}

The modified equation (\ref{N-J}) is:
\begin{eqnarray}\label{VinNJ}					
	\tau_3 =    
	\biggl[ 2\biggl\{ 1- \frac{R}{R+(1-f_{\mbox {II}})\Delta R_{\mbox I}   
	+ f_{\mbox {II}} \Delta R_{\mbox {II}}} 
	+ \frac{1}{2 \pi}   \sin \biggl( \frac{2 \pi R}{R+(1-f_{\mbox {II}})   
	 \Delta R_{\mbox I}+f_{\mbox {II}} \Delta R_{\mbox {II}}} \biggr)    
	 \biggr\} \biggr]^{-1}.
\end{eqnarray}	
							
Figure \ref{fig:tau3_fit} shows the fitted $\tau_3$ using Eq. (\ref{VinNJ}) 
as a function of temperature.
The fitting parameters are  $\Delta R_{\mbox {I}}$ and $\Delta R_{\mbox {II}}$.
The reduced $\chi^2$ is $1.7$. 
The modified Ps-bubble model represent the temperature 
dependence of $\tau_3$ well.

The fitted exuding depths are 
$\Delta R_{\mbox {I}} = 0.130\pm0.005$ nm and 
$\Delta R_{\mbox {II}} = 0.218\pm0.003$ nm.
In the Ps-bubble model, $\Delta R$ should correlate with the van der Waals 
radius of the surrounding atoms.
In fact, $\Delta R$, which is determined by fitting data of 
well-characterized small-pore materials such as zeolites for 
Eq. (\ref{N-J}), is 0.166 nm \cite{ujihira,Nakanishi88}, 
 which agrees with the 0.166 nm average of van der Waals radii 
 of atoms of SiO$_4$, the main component of zeolite.
 In our result, $\Delta R_{\mbox {I}}$ = 0.130 nm is close to the van 
der Waals radius  of oxygen (0.155 nm) and hydrogen (0.120 nm).
Contrarily, $\Delta R_{\mbox {II}}$ is too large for the van der Waals radius.
This indicates that the higher density state has a larger contribution 
to the pick-off process than their own depth of the electron wave function.
This constitutes evidence that the higher density state, which has 
a bending hydrogen bond \cite{Urquidi PRL 99},
is more active than the lower density state.

\begin{figure}								
\begin{center}									
	\includegraphics[width=8.0cm]{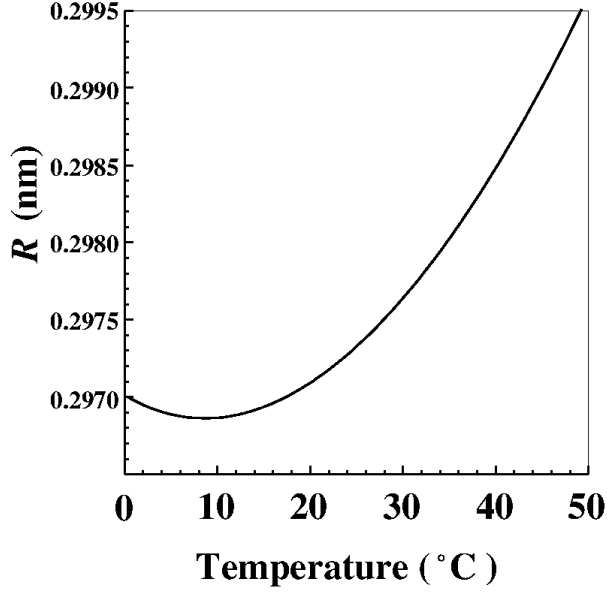}		
	\caption{\label{fig:radius}Temperature dependence of 
	  the Ps-bubble radius,
	 as calculated with $\gamma$ and fitted $\Delta R_{\mbox I}$ 
	 and $\Delta R_{\mbox {II}}$.}
\end{center}
\end{figure}

Next, the fitted $R$ supports the Ps-bubble idea.
As shown in Fig. \ref{fig:radius}, the radius of the Ps-bubble, $R$, based on 
our fit, is about 0.3 nm.
Liquid water has intrinsic vacancies in its structure while its structure 
is changing rapidly. 
Those vacancies produce a hexagonal structure having an oxygen atom at
each vertex connected by hydrogen bonds.
Considering the distance between the nearest O$\cdots$O, 0.28 nm \cite{Bosio83},
and the van der Waals radius of water, the vacancy is smaller than a Ps-bubble.
Therefore $o$-Ps should spread water molecules apart to exist in water.
This fact supports the need for a balance between the zero-point of $o$-Ps 
energy and the surface tension of water. 

In addition, our model shows that $R$ has a minimum at 8\doci.
The radius of Ps-bubble at 50\doci\ is 1.009 times the radius at 8\doci.
This size relationship is consistent with the fact that the distance 
between the nearest neighbors of O$\cdots$O pair at 50\doci\ is 1.011 
times the distance at 4\doci\ \cite{Narten67}. 

\section{Conclusion}
We precisely measured the temperature dependence of the long lifetime of 
a positron ($\tau_3$) in water at temperatures between 0 and 50\doci. 
The \tsan\ decreases smoothly as temperature rises; it is $1.839\pm0.015$ ns
 at 20\doci.
 This fact is explained by combining two models in which the water consists of 
two types of molecular bonds, and in which positroniums push those bonds
apart to form Ps-bubbles.
We also found that \tsan\ is sensitive to the electron states of the two 
types of molecular bonds.

\end{document}